\documentclass[aps,preprint,showpacs,preprintnumbers,nofootinbib,
               amsmath,amssymb,superscriptaddress]{revtex4}
\usepackage{graphicx}

\begin{document}

\title{Quark description of the Nambu-Goldstone bosons \\
 in the color-flavor locked phase}
\author{Kenji Fukushima}
\email{kenji@lns.mit.edu}
\affiliation{Center for Theoretical Physics, Massachusetts Institute
 of Technology, Cambridge, Massachusetts 02139}
\affiliation{Department of Physics, University of Tokyo,
 7-3-1 Hongo, Bunkyo-ku, Tokyo 113-0033, Japan}
\begin{abstract}
 We investigate the color-singlet order parameters and the quark
 description of the Nambu-Goldstone (NG) bosons in the color-flavor
 locked (CFL) phase. We put emphasis on the NG boson (phason) called
 ``H'' associated with the $\mathrm{U_B(1)}$ symmetry breaking. We
 qualitatively argue the nature of H as the second sound in the
 hydrodynamic regime. We articulate, based on a diquark picture, how
 the structural change of the condensates and the associated NG bosons
 occurs continuously from hadronic to CFL quark matter if the
 quark-hadron continuity is realized. We sharpen the qualitative
 difference between the flavor octet pions and the singlet phason. We
 propose a conjecture that superfluid H matter undergoes a crossover
 to a superconductor with tightly-bound diquarks, and then a crossover
 to superconducting matter with diquarks dissociated.
\end{abstract}
\preprint{MIT-CTP 3479}
\maketitle

%%%%%%%%%%   Introduction   %%%%%%%%%%

\section{introduction}
Quantum Chromodynamics (QCD) describing the dynamics of quarks and
gluons has the rich phase structure depending on the temperature and
the baryon density. In particular, the region at low temperature and
high density is the arena of a number of possibilities competing. Our
knowledge in condensed matter physics provides us with useful
information on dense QCD matter. For example, we know that the
attractive force between electrons mediated by phonons brings about
superconductivity of metals, and that a fermionic $^3\mathrm{He}$
liquid has the superfluid (B, A, and $\mathrm{A}_1$) phases
originating from pair condensation of $^3\mathrm{He}$
atoms~\cite{leg75,vol00}. It is natural to pursue similar
possibilities for nucleons~\cite{tam70} and quarks~\cite{bai84}.

Actually, the pairing force in the NN interaction leads to a finite
$^1S_0$ gap in nuclear matter. The superfluid properties of neutron
matter and neutron-rich nuclei have been paid much attention in both
nuclear and neutron star physics. When the baryon density is large
enough, a triplet-state $^3P_2$ superfluidity of neutron matter is
possibly favored~\cite{tam70,bed03}. At extremely high density, quarks
are expected to participate in the dynamics directly~\cite{col75} and
eventually the perturbative QCD calculation at weak coupling is
feasible owing to asymptotic freedom. As a matter of fact, the one
gluon exchange interaction between quarks is attractive in itself in
the color anti-triplet channel, which results in the
color-superconducting phase~\cite{son99,sch99_0}.

In the case of three massless flavors, the color-flavor locked (CFL)
phase~\cite{alf99} is well established at sufficiently high density.
The CFL phase preserving a symmetry under the color-flavor rotation is
similar to the B phase of $^3\mathrm{He}$ where a symmetry remain when
the orbit and spin rotations are locked together. Interestingly
enough, it has been discussed that the spontaneous breaking of global
symmetries in the CFL phase has the same pattern as that in hadronic
matter, as briefly reviewed below. It leads us to a conjecture that
superfluid (hyper-nuclear) hadronic matter can be smoothly connected
to CFL quark matter (i.e., quark-hadron continuity~\cite{sch99}). If
this happens, the physical content of the Nambu-Goldstone (NG) bosons
must continuously change from one to the other phase.

In order to investigate the low energy dynamics governed by the NG
bosons in the CFL phase, the non-linear representation of the chiral
effective Lagrangian has been studied~\cite{cas99,hon99,bed02,red03}.
There are, on the other hand, some attempts to describe the NG bosons
in terms of quarks directly~\cite{rho00,jai03}. An apparent problem
thus is that the effective Lagrangian approach suggests the (pionic)
NG bosons consisting of \textit{four} quarks, i.e., two quarks and two
quark-holes. Due to this structure, as discussed
in~\cite{cas99,hon99,bed02}, the inverse mass ordering of the NG
bosons occurs, which leads to the possible meson condensation in the
presence of finite quark masses or an electron chemical potential.
The fact that the NG bosons are made of four quarks has not been fully
taken into account in quark based calculations. [This has been
considered with respect to not the NG bosons but the sigma
meson~\cite{jai03}.] Hence, it is important to formulate the quark
description of the NG bosons in the CFL phase clearly.

The purpose of this paper is to give plain expressions for the NG
bosons in the CFL phase. Such a description enables us to have a clear
speculation on the quark-hadron continuity from the point of view of
the quark content of the condensates and the associated NG bosons. We
would emphasize that the structural change of the quark content can be
intuitively understood, even though it is hard to imagine how the
correspondence is realized, as originally conjectured in \cite{sch99},
between the color-flavor octet quarks and the flavor octet baryons, or
between the massive gauge bosons and the vector mesons.

%%%%%%%%%%   Nambu-Goldstone bosons   %%%%%%%%%%

%---   Chiral Symmetry Breaking   ---%

\section{Nambu-Goldstone bosons}
We shall limit our discussion to the chiral limit with three massless
flavors since we are interested in the ideal realization of the NG
bosons at first. Although a finite $m_s$ might change the phase
structure qualitatively~\cite{sch99_3}, the chiral limit would be a
well-defined starting point and an optimal underpinning to consider
more complicated situations.

In the superfluid phase of diquarks the relevant degrees of freedom
are given by diquarks rather than quarks because quarks are all
gapped. \textit{This is a dynamical assumption meaning that we are
working in the CFL phase.} The statement that the NG bosons in the CFL
phase consist of four quarks is generic not depending on the
assumption. The picture, as we argue below, that the four-quark boson
is regarded as a diquark-dihole state is derived from the assumption.
In other words we specify our problem by taking a diquark model
picture.

We define the left- and right-handed diquark fields as follows;
\begin{equation}
 \phi^\ast_{\text{L}i\alpha}=\epsilon_{ijk}
  \epsilon_{\alpha\beta\gamma}\bar{q}^{\text{C}}_{\text{L}j\beta}
  q_{\text{L}k\gamma}, \quad
 \phi^\ast_{\text{R}i\alpha}=\epsilon_{ijk}
  \epsilon_{\alpha\beta\gamma}\bar{q}^{\text{C}}_{\text{R}j\beta}
  q_{\text{R}k\gamma},
\end{equation}
where the complex conjugation ($\ast$) is attached in order that
$\phi_{\text{L,R}i\alpha}$ transforms as a triplet under the color and
flavor rotations as in \cite{cas99}. The Latin and Greek indices
represent flavor and color, respectively. It is assumed that dominant
diquarks are anti-symmetric in spin, color, and therefore flavor.
Since the condensate has positive parity, which is favored
energetically, we have
$\langle\phi_{\text{L}i\alpha}\rangle=-\langle\phi_{\text{R}i\alpha}
\rangle\propto\delta_{i\alpha}\Delta$ in the CFL phase if we choose a
certain gauge. The left-handed diquark condensate breaks the
symmetries as $\mathrm{SU_C}(3)\times\mathrm{SU_L}(3)\times
\mathrm{U_L}(1)\to\mathrm{SU_{C+L}}(3)\times\mathrm{Z}_{\text{L}}(2)$
($\phi_{\text{L}}$ is invariant under $q_{\text{L}}\to-q_{\text{L}}$).
The right-handed diquark condensate breaks the symmetries
likewise~\cite{alf99}. The residual symmetry is in a sense similar
to the custodial symmetry in the Higgs-Kibble model~\cite{kib67}.
After all, the chiral symmetry breaking takes place in the CFL phase
as
\begin{equation}
%\begin{split}
 \mathrm{SU_C}(3)\times\mathrm{SU_L}(3)\times\mathrm{SU_R}(3)\times
  \mathrm{U_V}(1)\times\mathrm{U_A}(1) % \\
 \to\mathrm{SU_{C+L+R}}(3)\times\mathrm{Z_L}(2)\times\mathrm{Z_R}(2),
\label{eq:sym_breaking}
%\end{split}
\end{equation}
as argued first in \cite{alf99}. We should remark that the
$\mathrm{U_A}(1)$ symmetry is explicitly broken due to the axial
anomaly and thus the residual discrete symmetry is only
$\mathrm{Z}(2)$ in fact. Nevertheless we leave both
$\mathrm{Z_{L,R}}(2)$ symmetries for later convenience. In the
subsequent paragraphs we will clarify the nature of the NG bosons
associated with (\ref{eq:sym_breaking}).

Let us first consider about the \textit{pions} in the CFL phase, that
is, the mesonic excitations forming an octet of
$\mathrm{SU_{C+L+R}}(3)$. As stated above, a vectorial symmetry is
preserved when the color rotation is locked with both the left- and
right-handed flavor (opposite) rotations. Only the axial part of
chiral symmetry (i.e., coset space
$\mathrm{SU_L}(3)\times\mathrm{SU_R}(3)/\mathrm{SU_{L+R}}(3)$) is
broken actually and the NG bosons appear corresponding to the axial
fluctuations. The important point is that the pions should be
colorless because the \textit{local} color symmetry is never broken
spontaneously~\cite{eli75}. In other words, gapless fluctuations with
non-trivial color charge are to be absorbed in the longitudinal
polarization of the gauge fields (Anderson-Higgs mechanism). As the
resulting NG bosons must be all colorless, it is rather preferable
to begin with color-singlet order parameters characterizing the
symmetry breaking pattern~(\ref{eq:sym_breaking}).

The simplest choice is to form a singlet order parameter from a
triplet $\phi_{\text{L,R}i\alpha}$ and an anti-triplet
$\phi_{\text{L,R}i\alpha}^\ast$ in color space. Then we have
$6\times6=36$ combinations (left- and right-handed triplet and
anti-triplet) for the flavor indices. The condition that it should
have positive parity and break chiral symmetry reduces those into $17$
combinations as follows;
\begin{align}
 \tilde{\sigma} &=\phi^\ast_{\text{L}i\alpha}\phi_{\text{R}i\alpha}+
  \phi^\ast_{\text{R}i\alpha}\phi_{\text{L}i\alpha},
\label{eq:op_meson} \\
 C^a_{\text{V}} &=\phi^\ast_{\text{L}i\alpha}\Bigl(
  \frac{\lambda^a}{2}\Bigr)_{ij}\phi_{\text{L}j\alpha}
  +\phi^\ast_{\text{R}i\alpha}\Bigl(\frac{\lambda^a}{2}\Bigr)_{ij}
  \phi_{\text{R}j\alpha}, \\
 C^a_{\text{A}} &=\phi^\ast_{\text{L}i\alpha}\Bigl(
  \frac{\lambda^a}{2}\Bigr)_{ij}\phi_{\text{R}j\alpha}
  +\phi^\ast_{\text{R}i\alpha}\Bigl(\frac{\lambda^a}{2}\Bigr)_{ij}
  \phi_{\text{L}j\alpha},
\end{align}
where $\lambda^a/2$ $(a=1,\dots,8)$ form an algebra of
$\mathrm{su}(3)$ in flavor space. $\tilde{\sigma}$ is a singlet under
the flavor $\mathrm{SU_V}(3)$ transformation, but neither
$C^a_{\text{V}}$ nor $C^a_{\text{A}}$ are. The essential point is that
\textit{only $\tilde{\sigma}$ serves as a proper order parameter for
the CFL phase.} This can be seen algebraically from
$\mathrm{tr}\lambda^a=0$. Otherwise ($\langle
C_{\text{V,A}}^a\rangle\neq0$), no vectorial subgroup of chiral
symmetry would survive. One might wonder where the color and flavor
are locked in the order parameter (\ref{eq:op_meson}) as they should
be. The answer is that the choice of $\tilde{\sigma}$ already reflects
the fact that the system is in the CFL phase. This is because the only
way to preserve the chiral $\mathrm{SU_V}(3)$ symmetry with a finite
diquark condensate is the color-flavor locking and thus the presence
of the unbroken $\mathrm{SU_V}(3)$ symmetry signifies the CFL phase
(see also the argument in \cite{hon03}). In the construction of the
chiral effective Lagrangian, the choice of the flavor-singlet currents
($X^\dagger\partial X$ and $Y^\dagger\partial Y$ in \cite{cas99})
corresponds to the choice of (\ref{eq:op_meson}) in the present
formulation.

As we stated before, the four-quark nature is generically derived from
the symmetry breaking pattern (\ref{eq:sym_breaking}) and the
color-singletness of the order parameter. The diquark description like
(\ref{eq:op_meson}) is, however, based on the diquark model picture.
Actually the four-quark order parameter can be given by other
expressions once some Fierz transformation is applied to
(\ref{eq:op_meson}). In principle one cannot distinguish the CFL phase
from the hadronic phase with exotic chiral symmetry
breaking~\cite{wat03} only from the symmetry breaking pattern.

Apparently $\tilde{\sigma}$ is a counterpart of the ordinary sigma
meson, $\sigma\sim\bar{q}_{\text{R}}q_{\text{L}}+\bar{q}_{\text{L}}
q_{\text{R}}$, which is regarded as a four-quark object as discussed
in \cite{jai03}. We must be cautious about this statement, however. It
does \textit{not} mean that $\tilde{\sigma}$ is a \textit{bound state}
of two quarks and two holes
 \footnote{For convenience, we will simply call a state excited by the
 operator $q$ a hole. It is actually mixed with an anti-quark whose
 contribution is negligible at high density and low temperature.}.
We should rather understand that one diquark (dihole) only supplies
the color charge as a background compensating for the other dihole
(diquark). Also it should be noted here that $\tilde{\sigma}$ has an
essential difference from $\sigma$. Under either $\mathrm{Z_L}(2)$ or
$\mathrm{Z_R}(2)$ transformation, $\tilde{\sigma}$ is invariant but
$\sigma$ is not. It should be noted that $\langle\sigma\rangle$ breaks
the $\mathrm{U_A}(1)$ symmetry up to $\mathrm{Z}(2)$ that amounts to
only a part of unbroken $\mathrm{U_V}(1)$. Therefore
$\mathrm{Z_L}(2)\times\mathrm{Z_R}(2)$ in the CFL phase is a larger
symmetry than $\langle\sigma\rangle$ leaves
 \footnote{When superfluidity occurs, the $\mathrm{U_V}(1)$ symmetry
 is also broken up to $\mathrm{Z}(2)$ in addition to
 $\mathrm{U_A}(1)\to\mathrm{Z}(2)$ due to the chiral condensate. The
 residual symmetry in this case is not
 $\mathrm{Z}(2)\times\mathrm{Z}(2)$ but only $\mathrm{Z}(2)$ and
 certainly smaller than $\mathrm{Z_L}(2)\times\mathrm{Z_R}(2)$
 realized in the CFL phase.}.
This difference is responsible for the inverse mass ordering of the
NG bosons~\cite{cas99} and, in general, the attribute of higher
dimensionful order parameters as discussed in \cite{wat03}.

Now that the order parameter is given by (\ref{eq:op_meson}), the NG
bosons can be read from the operator identity to define the
spontaneous symmetry breaking as formulated first in \cite{gol62}. Let
$Q_{\text{A}}^a$ be generators of the axial part of chiral symmetry,
i.e., $Q_{\text{A}}^a=\int\mathrm{d}x\,j_{\text{A}\mu=0}^a=
\int\mathrm{d}x\,\bar{q}\gamma_5\gamma_0(\lambda^a/2)q$. By using the
anti-commutation relation,
$\{\lambda^a/2,\lambda^b/2\}=\delta^{ab}/3+d^{abc}\lambda^c/2$, we can
readily show
\begin{equation}
 \bigl[\mathrm{i}Q_{\text{A}}^a, \tilde{\pi}^b\bigr]
  =\frac{\mathrm{i}}{3}\delta^{ab}\tilde{\sigma}
  +\mathrm{i}d^{abc}C_{\text{A}}^c
\label{eq:identity_ng}
\end{equation}
with
\begin{equation}
 \tilde{\pi}^a=\phi_{\text{L}i\alpha}^\ast\Bigl(\frac{\lambda^a}{2}
  \Bigr)_{ij}\phi_{\text{R}j\alpha}-\phi_{\text{R}i\alpha}^\ast\Bigl(
  \frac{\lambda^a}{2}\Bigr)_{ij}\phi_{\text{L}j\alpha}.
\end{equation}
In general, as explicitly seen in the
Umezawa-Kamefuchi--Lehmann-K\"{a}ll\'{e}n representation, we can prove
that $\tilde{\pi}^a$ couples to a massless state if the expectation
value of the right-hand-side of (\ref{eq:identity_ng}) is
non-vanishing \cite{gol62}. As a result, it follows that
$\tilde{\pi}^a\to Z^{1/2}\chi^a+\cdots$ and
$j_{\text{A}\mu}^a\to (f_{\tilde{\pi}}\partial_0\chi^a,\,
f_{\tilde{\pi}}v^2\partial_i\chi^a)+\cdots$ asymptotically, where $v$
is the velocity, $\chi^a$ is the (asymptotic) NG boson operator, and
$f_{\tilde{\pi}}Z^{1/2}=\langle\tilde{\sigma}/3\rangle$. This
non-perturbative relation may be made use of to determine a
non-perturbative value of $f_{\tilde{\pi}}$. [c.f.\ 
$f_\pi=\langle\sigma\rangle$ in the mean-field analysis of the linear
sigma model.] The four-quark operators, $\tilde{\pi}^a$, are thus the
interpolating fields of the pions in the CFL phase. Equivalently, by
using the Jacobi identity, one can intuitively regard the pions as
fluctuations around the order parameter, i.e.,
$\tilde{\pi}^a\sim[\mathrm{i}Q_{\text{A}}^a,\tilde{\sigma}]$.

A similar description in terms of four-quark states in terms of
diquarks (diquark--dihole or diquark--anti-diquark) has been argued in
diquark models \cite{shu03}. In contrast to the normal hadronic phase,
however, we would emphasize that the four-quark nature in the CFL
phase is \textit{rigid} and there is no mixing with the
quark--hole nor quark--anti-quark component. This is because of the
unbroken $\mathrm{Z_L(2)\times Z_R(2)}$ symmetry in the CFL phase. The
``parity'' under either $\mathrm{Z_L(2)}$ or $\mathrm{Z_R(2)}$
transformation becomes a good quantum number. The quark--hole or
quark--anti-quark object has odd ``parity'', while the diquark--dihole
or diquark--anti-diquark object has even ``parity''. Therefore, the
dominant Fock-state of the pionic excitation in the CFL phase is
considered as a four-quark state, apart from the instanton effect.

%--   Superfluidity   --%

Next let us consider about the \textit{phason}, that is, the NG boson
in connection with the superfluidity. We can construct color-singlet
order parameters with a non-vanishing baryon number. The simplest
choice is to form a color-singlet from three triplets. In this case
the flavor indices are more complicated than in the pion case, namely,
$6\times6\times6=216$ combinations. They can be reduced to 10
combinations by parity and anti-symmetric nature in color;
\begin{align}
 H &= H_{\text{L}}-H_{\text{R}} =
 \epsilon_{ijk}\epsilon_{\alpha\beta\gamma}\bigl(
  \phi_{\text{L}i\alpha}\phi_{\text{L}j\beta}\phi_{\text{L}k\gamma}
  -\phi_{\text{R}i\alpha}\phi_{\text{R}j\beta}\phi_{\text{R}k\gamma}
  \bigr),
\label{eq:op_h} \\
 \tilde{H}_1 &=\epsilon_{ijk}\epsilon_{\alpha\beta\gamma}\bigl(
  \phi_{\text{L}i\alpha}\phi_{\text{L}j\beta}\phi_{\text{R}k\gamma}
  -\phi_{\text{R}i\alpha}\phi_{\text{R}j\beta}\phi_{\text{L}k\gamma}
  \bigr), \\
 \tilde{H}_8^a &=\epsilon_{\alpha\beta\gamma}\Bigl\{\epsilon_{ijk}
  \phi_{\text{L}j\alpha}\phi_{\text{L}k\beta}\Bigl(\frac{\lambda^a}{2}
  \Bigr)_{il}\phi_{\text{R}l\gamma}
   -\epsilon_{ijk}\phi_{\text{R}j\alpha}
  \phi_{\text{R}k\beta}\Bigl(\frac{\lambda^a}{2}\Bigr)_{il}
  \phi_{\text{L}l\gamma}\Bigr\}.
\end{align}
In our notation the above expression of $H$ gives positive parity
since $\phi_{\text{L,R}}\to-\phi_{\text{R,L}}$ under the parity
transformation. In the CFL phase both $H$ and $\tilde{H}_1$ are
$\mathrm{SU_V}(3)$-singlets and can take a finite expectation value,
while $\langle\tilde{H}_8^a\rangle=0$. A non-vanishing expectation
value of $H$ would break only the $\mathrm{U_V}(1)$ and
$\mathrm{U_A}(1)$ symmetries up to
$\mathrm{Z_L}(2)\times\mathrm{Z_R}(2)$. On the other hand,
$\tilde{H}_1$ is variant also under the axial part of chiral symmetry
as well as the $\mathrm{U}(1)$ symmetries. A natural interpretation on
$\tilde{H}_1$ is a composite state made of $\tilde{\sigma}$ and $H$.
Therefore we will give little thought about $\tilde{H}_1$ hereafter
though $\langle\tilde{H}_1\rangle$ could be qualified as an order
parameter. We would mention that one can directly prove that
$(\mathrm{Tr}X^\dagger\partial X)^2+(\mathrm{Tr}Y^\dagger\partial Y)^2$
in \cite{cas99} originates from
$(\partial H_{\text{L}})^2+(\partial H_{\text{R}})^2$ in the present
formulation.

The interpolating field of the NG boson can be deduced in the same way
as in the pion case. The generator of the $\mathrm{U_V}(1)$ rotation
is given by $Q_{\text{V}}=\int\mathrm{d}x\,j_{\mu=0}
=\int\mathrm{d}x\,\bar{q}\gamma_0 q$. Then, as in the familiar
Goldstone model \cite{gol61}, the phase of the order parameter is the
NG boson (phason) \cite{and66}. We can easily show
\begin{equation}
 [\mathrm{i}Q_{\text{V}},H] = -6\mathrm{i}H.
\label{eq:com_h}
\end{equation}

Interestingly enough, the $H$ field defined in (\ref{eq:op_h}) has the
same quark content as the H-dibaryon~\cite{jaf77} as already pointed
out in \cite{sch99,cas99,pis00}. Let us clarify physical interpretation on
the $H$ field. The $H$ field has two degrees of freedom. In the
$\mathrm{U_V}(1)$ broken phase the vacuum state is described by the
coherent state of $H$, which is the superposition of many $H$ and
$H^\ast$ states. [$H$ is a complex scalar field and $H^\ast$ is its
conjugate.] If we choose the real part, $(H+H^\ast)/2$, as a
condensate, the phason, $\chi_H$, is given by the imaginary part,
$\mathrm{i}(H-H^\ast)/2$. The relation between the real and imaginary
parts of $H$ under the $\mathrm{U_V}(1)$ transformation can be
understood in analogy with the $\mathrm{U_A}(1)$ rotation for
$\sigma\sim\bar{q}_{\text{L}}q_{\text{R}}
+\bar{q}_{\text{R}}q_{\text{L}}$ and
$\eta_0\sim\bar{q}_{\text{L}}q_{\text{R}}
-\bar{q}_{\text{R}}q_{\text{L}}$. It would be legitimate to regard
both parts of $H$ as \textit{particles} just like $\sigma$ and
$\eta_0$ (which eventually becomes $\eta'$ due to mixing with $\eta_8$
when $m_{\text{s}}>m_{\text{u}}\simeq m_{\text{d}}$), though there is
no clear distinction for the $H$ field.

The velocity of $\chi_H$ has been calculated at zero
temperature~\cite{cas99} and the zero temperature value, $v^2=1/3$,
has been prevailing in model studies~\cite{red03}. When it comes to the
velocity of $\chi_H$ at $T\neq0$, it is important to distinguish the
collisionless regime from the hydrodynamic one. Let $\omega$ and
$\tau$ be the frequency of $\chi_H$ and the characteristic collision
time which is roughly of order $1/T$. The zero temperature estimate is
expected to work in the collisionless regime, $\omega\tau\gg 1$,
meaning H with energies much higher than $T$. The collisionless regime
shrinks with increasing temperature. Then, the hydrodynamic
description becomes more relevant in the hydrodynamic regime,
$\omega\tau\ll 1$.

One might consider that $\chi_H$ is not a propagating (particle) mode
but a diffusive mode in the hydrodynamic regime, for it looks like a
density fluctuation. This is not the case in the superfluid phase
since the number density is no longer conserved. Actually the order
parameter field and the entropy density are mixed to lead to a
propagating mode (second sound) in a superfluid~\cite{kaw70,for75}. In
a dilute superfluid system the nature of hydrodynamic modes have been
investigated~\cite{for75,gri93,lee59}. Then in the hydrodynamic regime
it has been actually shown that the phason mainly corresponds to the
second sound in the hydrodynamic regime. This consequence makes sense;
the second sound only exists in the superfluid (symmetry-broken)
phase, while the first sound is present in both the superfluid and
normal phases. As to the QCD context, due to the asymptotic freedom at
high density, we can expect that the analyses in a dilute system are
relevant to our problem. Once the $\chi_H$ propagation is identified
as the second sound, the velocity of $\chi_H$ in the hydrodynamic
regime would be quite different from the zero temperature value.

In general a superfluid at finite temperature has a non-vanishing
normal component whose density is $\rho_{\text{n}}$ as well as a
superfluid one whose density is $\rho_{\text{s}}$. The two-fluid
dynamics gives the general expression for the second sound speed as
\begin{equation}
 c_2 = \sqrt{\frac{\rho_{\text{s}}s^2 T}{\rho_{\text{n}}\rho c_v}},
\end{equation}
where $\rho=\rho_{\text{n}}+\rho_{\text{s}}$. The entropy density and
the specific heat are denoted by $s$ and $c_v$ respectively. The
velocity, which is proportional to $\sqrt{\rho_{\text{s}}}$,
approaches zero as the system goes closer to the critical temperature.
This observation is analogous to the pion velocity vanishing at the
critical temperature~\cite{son02_2} though the physics is different.

Whether we should consider in the collisionless or hydrodynamic regime
depends on the energy of H and the temperature. The energy and
temperature dependence of the velocity of $\chi_H$ will be of
importance, for example, in $\chi_H$-involved processes in the
proto-neutron star at temperatures of tens MeV~\cite{red03}.

%%%%%%%%%%   Structural change   %%%%%%%%%%

\section{structural change}
Now we shall consider how the quark-hadron continuity can be viewed
from the quark content of the color-singlet condensates and the NG
bosons. We will find it important to recognize that the meaning of
continuity is twofold; one is with respect to chiral symmetry and
the other is related to the confinement-Higgs crossover~\cite{fra79}.
Their physical meanings are distinct.

%---   Mesonic Excitations   ---%

For the moment we will adopt a working definition; ``hadronic matter''
is used for the phase with $\langle\sigma\rangle\neq0$, and ``CFL
phase'' for the phase with $\langle\sigma\rangle=0$. This is in
accordance with the prevailing convention in the chiral Lagrangian
approach~\cite{cas99,pis00}. In hadronic matter pions are mainly
composed of a quark ($q$) and an anti-quark ($\bar{q}$) in the
constituent quark picture.

As discussed some time ago~\cite{jaf77_2} there can be a significant
mixture of $\bar{q}^2 q^2$ even in the hadronic phase and we have not
only $\langle\sigma\rangle\neq0$ but also
$\langle\tilde{\sigma}\rangle\neq0$. In the CFL phase, i.e., in the
phase having the $\mathrm{Z_L}(2)\times\mathrm{Z_R}(2)$ symmetry
(apart from the axial anomaly), $\langle\sigma\rangle$ becomes zero
and any $\bar{q}q$ component in the pions vanishes. The quark-hadron
continuity is realized by a non-vanishing common ingredient of the
pions, $\tilde{\pi}^a$, made of a diquark and a dihole. The instanton
effect, of course, breaks the $\mathrm{U_A}(1)$ symmetry explicitly
and blurs a sharp phase separation. The important point is, however,
that \textit{we can have a definite limit where hadronic matter is
distinguished from the CFL phase with respect to the pions}
 \footnote{The limit where $\mathrm{Z_L}(2)\times\mathrm{Z_R}(2)$ is
 present does not corresponds to the \textit{absence} of instantons.
 Otherwise, not only $\mathrm{U_A}(1)$ but also chiral symmetry is
 not broken. This limit should be understood in the same sense as the
 \textit{effective} restoration of the $\mathrm{U_A}(1)$ symmetry
 discussed in \cite{shu94}.}.

%---   Superfluid Component   ---%

\begin{figure}
\begin{center}
 \includegraphics[width=10cm]{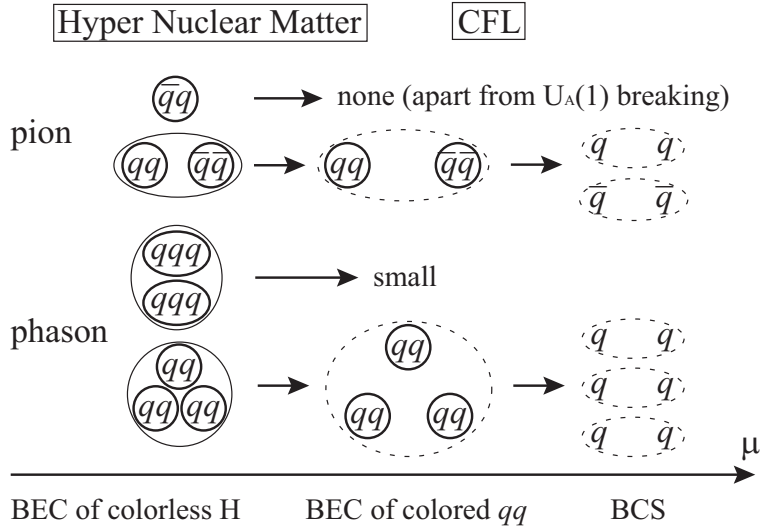}
 \caption{Schematic picture of the structural change and the two-step
 crossover.}
\label{fig:sketch}
\end{center}
\end{figure}

In contrast, there remain more or less controversial points in the
discussion about superfluidity. This is because hyper-nuclear matter
has not been completely understood. It is expected at least that
hyper-nuclear matter has a finite gap leading to a superfluid phase
where $\mathrm{U_V}(1)$ is broken into $\mathrm{Z}(2)$
 \footnote{As stated in \cite{son02}, the $\mathrm{U_B}(1)$ symmetry
 generated by the baryon charge is different from $\mathrm{U_V}(1)$
 discussed here. Since we always deal with color-singlet objects,
 however, the $\mathrm{U_V}(1)$ rotation is equivalent with three
 (six, nine, etc) times $\mathrm{U_B}(1)$ rotation, apart from the
 discrete symmetries. In this sense, though it is not rigorously
 correct, we will use a sloppy notation and sometimes write
 $\mathrm{U_V}(1)$ and $\mathrm{Z_V}(2)$ to mean
 $\mathrm{U_B}(1)$ and $\mathrm{Z_B}(2)$ respectively.}.
There are 8 baryons forming an octet in the $\mathrm{SU_V}(3)$
symmetry. The superfluid condensate is an $\mathrm{SU_V}(3)$ singlet,
which can be constructed by two baryon octets as
\begin{equation}
 \Delta_{NN}\sim\bigl\langle (\Sigma^0)^2 +(\Lambda^0)^2 +\Sigma^+
  \Sigma^- +\Sigma^- \Sigma^+
  +p\Xi^- +\Xi^- p +n\Xi^0 +\Xi^0 n \bigr\rangle.
\end{equation}
Moreover we can also expect $\langle H\rangle\neq0$ ($H$ is given by
(\ref{eq:op_h}), i.e., a three diquark-like object) because it is not
prohibited by the symmetry. As a matter of fact, the H-dibaryon state
could be an admixture of two-nucleon-like and three-diquark-like
configurations. This can be typically seen in the different
treatments, namely, the resonating group method (RGM)~\cite{oka80} and
the MIT bag model~\cite{jaf77}. Unlike the pion case, we have no
definite limit where we can give a rigorous statement about the
structural change of the superfluid component. We can only draw a
qualitative picture in an intuitive picture. In the phase where the
$\mathrm{Z_L}(2)\times\mathrm{Z_R}(2)$ symmetry is manifested, any
nucleon made of 3 quarks cannot be present, though a nucleon
\textit{pair} can be. Thus it is likely that the nucleon-like
condensate, $\Delta_{NN}$, becomes less significant in the CFL
phase. It means that the overlap of the wave-function in the
condensate becomes larger for $\langle H\rangle$ (probably deconfined
three-diquark condensate) then. This can be interpreted as a crossover
from a confined phase to a Higgs phase. [Actually the major difference
between the RGM and the MIT bag model comes from the way to impose the
confinement condition. The RGM is stricter in confinement than the
MIT bag model~\cite{oka80}.] We would note that a quite similar
picture (continuity between hyper-nuclear matter and partially
deconfined quark matter) has already been proposed some time ago in
the context of H-matter in the neutron star~\cite{tam91}. In contrast
to the pion case, the important message here is that \textit{the
continuity with respect to the phason is inherently
indistinguishable}. We summarized our point of view in
Table~\ref{tab:sum} and Fig.~\ref{fig:sketch} schematically.

\begin{table}
 \begin{tabular}{c|cc}
        & Hyper Nuclear Matter & CFL \\ \hline
 Chiral & $\langle\sigma\rangle\sim\langle\bar{q}q\rangle\neq0$
        & $\langle\sigma\rangle=0$ \\
 Symmetry  & $\langle\tilde{\sigma}\rangle\sim
  \langle\bar{q}^2 q^2\rangle\sim\mbox{small}$
        & $\langle\tilde{\sigma}\rangle\neq0$ \\
        & $\pi\sim \bar{q}q+(\bar{q}\bar{q})(qq)+\cdots$
        & $\tilde{\pi}\sim (\bar{q}\bar{q})(qq)+\cdots$ \\
        & \multicolumn{2}{c}{$\mathrm{Z}(2)\longrightarrow
            \mathrm{Z_V}(2)\mbox{ (instanton)}\longleftarrow
            \mathrm{Z_L}(2)\times \mathrm{Z_R}(2)$} \\ \hline
 Super- & $\Delta_{NN}=\langle\Sigma^2+\Lambda^2+N\Xi\rangle\neq0$
        & $\Delta_{NN}\sim\mbox{small}$ \\
 -fluidity & $\langle H\rangle\neq0$
        & $\langle H\rangle\neq0$ \\
        & $\chi_H\sim\mbox{2 nucleons}+\mbox{3 diquarks}$
        & $\chi_H\sim\mbox{3 diquarks}$ \\
        & \multicolumn{2}{c}{confinement--Higgs crossover} \\ \hline
 \end{tabular}
 \caption{Summary of qualitative changes from nuclear matter to CFL
 quark matter.}
\label{tab:sum}
\end{table}

%---   BEC - BCS of Diquarks   ---%

In principle the color-superconductivity can be a different phenomenon
from the chiral phase transition, though we have considered chiral
symmetry to characterize the phase. The situation is even similar to the
relation between the deconfinement and chiral phase transitions at
finite temperature \cite{fuk03}. There, the deconfinement transition
is a crossover in the presence of dynamical quarks in the fundamental
representation, while the chiral phase transition is well-defined in
the chiral limit. In the present case, the liberation of color degrees
of freedom corresponds to a crossover from hadronic matter to the
color-superconducting phase where colored diquarks play an essential
role. From the point of view of the NG bosons this can be seen as
dissociation of the hadrons into constituent diquarks. As far as $H$
is concerned, we can say in the following way; the hadronic phase has
a Bose-Einstein condensate (BEC) of the color-singlet H-dibaryon,
while the dissociated colored diquarks lead to a superconducting state
at higher baryon density, and yet they compensate for their color
charge to be a color-singlet in the CFL phase. In this sense, the
attractive force between diquarks controls the state of matter. If the
interaction is strong enough, the state is BEC-like, and otherwise, it
is BCS-like. This BEC-BCS crossover looks quite different from that
demonstrated in the $^3\mathrm{He}$ superfluid~\cite{leg80}, discussed
in the system at finite isospin chemical potential~\cite{son01}, and
investigated in the color-superconductivity~\cite{abu02}. In the
present case, the BEC-BCS crossover is seen, not in the Cooper pair
itself, but in the combination of two (for the pion) or three (for the
phason) Cooper pairs. This is actually a crossover from the BEC of H
to the BEC of diquarks.

So far in our discussion we have implicitly assumed the existence of
compact (tightly bound) diquarks. In the weak coupling regime diquarks
become spatially spreading, which would cause another (and more
familiar) crossover from the BEC of diquarks to the BCS. This type of
the BEC-BCS crossover is well-known and have been intensely studied in
condensed matter physics. Thus, as the baryon density increases, the
system should undergo the two-step crossover; BEC of colorless H $\to$
color-superconductivity, and then BEC of colored diquarks $\to$ BCS.\ \
This scenario in sketched in Fig.~\ref{fig:sketch}.

%%%%%%%%%%   Conclusions and an Outlook   %%%%%%%%%%

\section{conclusions and an outlook}
In this paper, we discussed the quark description of the NG bosons in
the CFL phase. We emphasized the four-quark nature of the pions
stemming from the residual Z(2) symmetry and the color-singletness. We
wrote down the plain expressions for the interpolating fields of the
NG bosons motivated by the diquark picture. Then we discussed the H
particle, that is the NG boson associated with the $\mathrm{U_B}(1)$
symmetry breaking, in the hydrodynamic regime. We point out the
velocity of the phason is given by the second sound speed which
vanishes at the critical temperature. Finally we proposed a conjecture
about the quark-hadron continuity. If the quark-hadron continuity is
realized, the NG boson must change its nature continuously from one to
the other phase. Based on the constituent diquark picture and the
quark-hadron continuity hypothesis, we drew the two-step crossover
scenario; BEC of H $\to$ BEC of diquarks, and then BEC of diquarks
$\to$ BCS.\ \ This scenario can be confirmed by developing the H
matter description discussed in \cite{tam91} or the tightly-bound
diquark picture as discussed in \cite{shu03}.

Although the diquark model leading to too light H dibaryons may not be
realistic in the vacuum, we can expect that it is a proper description
at high baryon density where the color-superconductivity occurs. A
three-body problem in terms of diquarks at finite density would be the
next step to go further into quantitative investigations. As shown in
Fig.~1 in \cite{shu03} the diquark interaction through the 't~Hooft
term is affected by the chiral condensate. This may cause an
entanglement between the chiral condensate and the diquark, in other
words, between the pions and the phason. It would be intriguing to
study how far the phase transition with respect to the pions can bring
about the structural change of the phason.

In this work the explicit breaking of the $\mathrm{U_A}(1)$ symmetry
has been regarded as an external perturbation smearing a sharp
distinction based on chiral symmetry. It would be a challenging
problem to study its effect not only on the pions but also on the
superfluid structure. Since it is known that a strong three-body
repulsion induced by the instanton effect makes the H-dibaryon weakly
bound or unbound theoretically~\cite{tak91}, the instanton-induced
interaction will affect the content of the superfluid component.

Our speculation implies that diquarks would become important for
the hadronic phase if it is close to CFL quark matter at high baryon
density. This must be a robust consequence even in the presence of
finite $m_{\text{s}}$ as long as there is no first order phase
transition. The nature of diquarks in the hadronic phase deserves
further investigation not only in the vacuum~\cite{ans93} but rather
at high baryon density. This is partially because the importance of
diquarks would be seen once the chiral condensate is vanishingly
small. A possible suggestion is that, near the chiral phase transition
at high temperature where the chiral condensate melts, the diquark
correlation can be relevant to thermodynamic quantities, as has been
already pointed out~\cite{shu04}.

In order to make our argument applicable, for instance, to neutron
star physics, it is necessary to take account of finite
$m_{\text{s}}$ and the neutrality conditions. The CFL phase is robust
as long as $m_{\text{s}}^2/\mu<2\Delta_{\text{CFL}}$~\cite{raj01}. As
discussed in \cite{sch99_3}, if $m_{\text{s}}$ is large enough to
suppress the hyperon number density, the transition from ordinary
nuclear matter to the CFL phase is discontinuous. Detailed
calculations, however, suggest the presence of H matter in the neutron
star~\cite{tam91}, though further analyses are needed to reveal the
actual properties of H matter in the cores of compact stellar objects.
Since the diquark picture tends to give rise to tightly-bound
H-dibaryon, we may well think that the calculation with diquarks taken
into account would result in the existence of H matter and our
discussion here is not altered qualitatively. If our scenario is
realized, then it would be quite interesting to see how the nature of
H (or $\chi_H$) affects the internal structure of the vortices in a
superfluid along the lines of \cite{iid02}.

We believe that our pictorial understanding could shed light on the
non-perturbative region between hadronic and quark matter and cast
novel and challenging problems.

\acknowledgments
The author thanks K.~Rajagopal for critical comments. He is also
grateful to M.~M.~Forbes, E.~Gubankova and O.~Schr\"{o}der for
conversation. He would like to acknowledge T.~Kunihiro and
M.~A.~Stephanov for useful comments. This work was partially supported
by Japan Society for the Promotion of Science for Young Scientists and
the U.S.\ Department of Energy (D.O.E.) under cooperative research
agreement \#DF-FC02-94ER40818.

\end{document}